\documentclass{article}
\usepackage{latexsym}

\begin{document}

\title{Staticity Theorem for Non--Rotating Black Holes with
Non--Minimally Coupled Self--Interacting Scalar Fields}

\author{Eloy Ay\'{o}n--Beato\\~\\
Departamento~de~F\'{\i}sica\\
Centro~de~Investigaci\'on~y~de~Estudios~Avanzados~del~IPN\\
Apdo.~Postal~14--740,~C.P.~07000,~M\'{e}xico,~D.F.,~MEXICO.}
\maketitle

\begin{abstract}
Self--interacting scalar field configurations which are
non--minimally coupled ($\zeta\neq0$) to the gravity of a strictly
stationary black hole with non--rotating horizon are studied. It
is concluded that for analytical configurations the corresponding
domain of outer communications is static.
\end{abstract}

\section{Introduction}

All the available ``no--hair'' theorems for non--rotating
stationary black holes are based in a staticity hypothesis
(see~\cite{Heusler96,Bekens96,Heusler98,Bekens98,Carter97} for
recent revisions on the subject), i.e., that the asymptotically
timelike Killing field {\boldmath$k$}, coinciding at the horizon
${\cal{H}}^{+}$ with its null generator, must be hypersurface
orthogonal in all the domain of outer communications $\langle
\!\langle \hbox{${\cal J}$\kern -.645em
{\raise.57ex\hbox{$\scriptscriptstyle(\ $}}}\rangle \!\rangle$. In
this way, in order to establish the uniqueness of the final state
of the gravitational collapse of all these systems it is needed to
make use of {\em staticity theorems}.

Generalizing a previous result by Lichnerowicz for space--times
without horizons~\cite{Lichne}, the staticity theorem
corresponding to vacuum black holes was proved by
Hawking~\cite{Hawking72a,H-E}, assuming the existence of strict
stationarity, i.e., that the Killing field {\boldmath$k$} is not
only timelike
($V\equiv-(\mbox{\boldmath$k$}|\mbox{\boldmath$k$})\geq0$) at
infinity, but in all the domain of outer communications $\langle
\!\langle \hbox{${\cal J}$\kern -.645em
{\raise.57ex\hbox{$\scriptscriptstyle(\ $}}}\rangle \!\rangle$.

The extension of this proof to the case of electrovac black holes
was proposed by Carter~\cite{Carter87}, and holds only assuming a
condition more restrictive than that one of strict stationarity.
This condition occurs to be unphysical, since it is violated even
for the black holes of the Reissner--Nordstr\"om family when they
have electric charges in the interval $4M/5<Q<M$.

Using a Hamiltonian approach, both staticity theorems, for the
vacuum and electrovac black holes, have been proved by Sudarsky
and Wald~\cite{SudarskyWald92,SudarskyWald93,Wald93} without the
previous restrictive hypotheses and using only a maximal slice,
the existence of which was proved later by Chru\'sciel and
Wald~\cite{ChrWald94}. More recently, this last result has been
extended, using the same technics, to the axi--dilaton gravity
coupled to electromagnetism which is derived from string theory to
low energies~\cite{Rogatko98}.

For minimally coupled scalar models, staticity theorems has been
also proved by Heusler under the strict stationarity
hypothesis~\cite{Heusler93,HeuStrau}. In this paper we will extend
the results by Heusler to the case of non--minimal coupling,
assuming again strict stationarity and also analyticity of the
scalar fields, establishing in this way that a non--rotating
strictly stationary black hole, corresponding to a
self--interacting non--minimally coupled scalar field, is static.

For minimally coupled scalar fields there is no need to impose the
existence of analyticity for static configurations, since the
elliptical nature of the corresponding Einstein--Scalar system
guarantees that all the fields are analytical in appropriate
coordinates (see the remarks of~\cite{Muller70a} about the
non--vacuum case). The situation is rather different for
non--minimally coupled systems, in this case the relevant
equations are not necessarily elliptic, and consequently the
existence of analyticity must be imposed as a supplementary
assumption (see the related discussion in~\cite{Zannias98}).
However, the analyticity hypothesis cannot be considered a too
strong one in the study of stationary black holes, since the
classification of them rests implicitly in this condition through
the Hawking strong rigidity theorem~\cite{Hawking72a,H-E}. This
theorem establishes that the event horizon of a stationary black
hole is a Killing horizon, i.e., there exist a Killing field
coinciding at the event horizon with the null generators of it
(for non--rotating horizons it is the same that the stationary
Killing field {\boldmath$k$}, but for rotating ones it is
different). In order to prove this celebrated theorem, Hawking
recurred to the existence of analyticity on all the fields
constituting the stationary black hole configuration.

Recent attempts to replace this condition by the more natural one
of smoothness have given only positive results for the region
interior to the event horizon \cite{FriedRW99}, hence they are of
little use in the subject of classification of stationary
black--hole exteriors.

The result we will establish is of great utility to eliminate the
staticity supposition from the ``no--hair'' theorems which has
been proved for non--minimally coupled scalar fields in presence
of non--rotating stationary black holes, see e.g.~\cite{Zannias95}
for the conformal case,
and~\cite{Saa96a,MayoBeke96,Saa96b,Bekens96} for other results
recently derived for more general coupling, where it is not only
assumed staticity but also spherical symmetry.

\section{The Staticity Theorem for Non--Minimally Coupled Scalar
Fields}

Let us consider the action for a self--interacting scalar field
non--minimally coupled to gravity
\begin{equation}
{\cal S}=\frac 12\int dv\left(\frac 1\kappa R
-(\nabla _\mu \Phi
\nabla ^\mu \Phi +U(\Phi ))-\zeta \,R\,\Phi ^2\right),
\label{eq:ac}
\end{equation}
where $\zeta$ is a real parameter (the values $\zeta=0$ and
$\zeta=1/6$ correspond to minimal and conformal coupling,
respectively).

The variations of this action with respect to the metric and the
scalar field, respectively, give rise to the Einstein equations
\begin{equation}
\left(1-\kappa\,\zeta\,\Phi^2\right)R_\nu^{~\mu}=
\kappa\left(\nabla_\nu\Phi\nabla^\mu\Phi
+\frac12\delta_\nu^{~\mu}\left(U(\Phi)-\zeta\,\Box\Phi^2\right)
- \zeta\,\nabla_\nu\nabla^\mu\Phi^2\right),
\label{eq:Ein}
\end{equation}
and the nonlinear Klein--Gordon equations
\begin{equation}
\Box\Phi-\frac12\frac{d\,U(\Phi)}{d\,\Phi}-\zeta\,R\,\Phi=0.
\label{eq:K-G}
\end{equation}

From this system of equations it must follows the existence of
staticity in the case of strictly stationary black holes with a
non--rotating horizon. As it was quoted at the beginning,
staticity means that the stationary Killing field {\boldmath$k$}
is hypersurface orthogonal, which is equivalent, by the Frobenius
theorem, to the vanishing of the twist 1--form
\begin{equation}
\omega_\alpha\equiv\frac12
\eta_{\alpha\beta\mu\nu}k^\beta\nabla^\mu{k}^\nu ,
\label{eq:tor}
\end{equation}
where {\boldmath$\eta$} is the volume 4--form.

In order to exhibit the existence of staticity we will find the
explicit dependence of {\boldmath$\omega$} in terms of $\Phi$, by
solving the following differential equations which must be
satisfied by the twist 1--form~\cite{Carter87}
\begin{equation}
\eta^{\mu\nu\alpha\beta}\nabla_\alpha\omega_\beta=
2k^{[\mu}{\cal{R}}^{\nu]},
\label{eq:rela}
\end{equation}
where
\[
{\cal{R}}^\mu\equiv{k^\nu}R_\nu^{~\mu},
\]
is the Ricci vector, that can be evaluated from Einstein equations
(\ref{eq:Ein}). Using the stationarity of the scalar field
\[
\mbox{\boldmath${\pounds}_k$}(\Phi)\equiv{k^\nu\nabla_\nu\Phi}=0,
\]
and the identity
\[
k^\nu\nabla_\nu\nabla^\mu\Phi^2=-\nabla^\mu{k}^\nu\nabla_\nu\Phi^2,
\]
the Ricci vector can be written as
\begin{equation}
\left(1-\kappa\,\zeta\,\Phi^2\right){\cal R}^\mu =
\kappa\left(\frac12k^\mu\left(U(\Phi)-\zeta\,\Box\Phi^2\right)
+ \zeta\,\nabla^\mu{k}^\nu\nabla_\nu\Phi^2\right).
\label{eq:flu}
\end{equation}

Replacing this expression in (\ref{eq:rela}),
using the Killing vector definition
\[
\nabla^\mu{k}^\nu=\nabla^{[\mu}k^{\nu]},
\]
it is obtained the following identity
\begin{equation}
\left(1-\kappa\,\zeta\,\Phi^2\right)\eta^{\mu\nu\alpha\beta}
\nabla_\alpha\omega_\beta =
2\,\kappa\,\zeta\,
k^{[\mu}\nabla^\nu{k}^{\alpha]}\nabla_\alpha\Phi^2,
\label{eq:eint}
\end{equation}
which, taking into account definition (\ref{eq:tor}), can be
rewritten as
\begin{equation}
\left(1-\kappa\,\zeta\,\Phi^2\right)\eta^{\mu\nu\alpha\beta}
\nabla_\alpha\omega_\beta =
\frac{2}{3}\kappa\,\zeta\,\eta^{\mu\nu\alpha\beta}
\nabla_\alpha\Phi^2\,\omega_\beta ,  \label{eq:ante}
\end{equation}
or equivalently, in the language of differential forms as follows
\begin{equation}
\mbox{\boldmath$d\omega$}=\mbox{\boldmath$d$}
\left(\ln\left[\left(1-\kappa\,\zeta\,\Phi^2\right)^{-2/3}
\right]\right)
\wedge\mbox{\boldmath$\omega$},  \label{eq:diform}
\end{equation}
where obviously the above expression is valid only in the regions
where $\Phi^2\neq1/\kappa\,\zeta$. The expression
(\ref{eq:diform}) can be written also as
\begin{equation}
\mbox{\boldmath$d$}\left(
\left(1-\kappa\,\zeta\,\Phi^2\right)^{2/3}\mbox{\boldmath$\omega$}
                   \right)=0,
\label{eq:closed}
\end{equation}
where it is understood once more that the equality is valid only
when $\Phi^2\neq1/\kappa\,\zeta$.

We will analyze now the value of the left hand side of
(\ref{eq:closed}) in the regions where $\Phi^2=1/\kappa\,\zeta$,
situation which is only possible for positive values of the
non--minimal coupling ($\zeta>0$). We claim that in the analytic
case these regions are composed from a countable union of lower
dimensional surfaces, hence, by the continuity of the left hand
side of (\ref{eq:closed}), the involved 1--form is also closed in
regions where $\Phi^2=1/\kappa\,\zeta$.

Lets examine the argument in detail. As it was mentioned at the
introductory Section, we suppose that the scalar field $\Phi$ and
the metric {\boldmath$g$} are analytical in appropriated
coordinates. First it must be noticed that
$\Phi^2\not\equiv1/{\kappa\,\zeta}$ in the whole of
$\langle\!\langle \hbox{${\cal J}$\kern -.645em
{\raise.57ex\hbox{$\scriptscriptstyle(\ $}}}\rangle \!\rangle $,
i.e., that the square of the scalar field does not take the value
$1/{\kappa\,\zeta}$ in every point of the domain of outer
communications. This is based in that the converse is in
contradiction with the fact that the asymptotic value of the
effective gravitational constant
\[
G_{\rm{eff}}\equiv\frac{G}{\left(1-\kappa\,\zeta\,\Phi^2\right)},
\]
must be positive and finite due to the know attractive character
of gravity at the asymptotic regions~\cite{MayoBeke96}. Since
$\Phi^2\not\equiv1/{\kappa\,\zeta}$, it can be shown (see
Ref.~\cite{MullerHRS73}) from the analyticity of $\Phi$, that if
the inverse images of the real values
$\left\{\pm1/\sqrt{\kappa\,\zeta}\right\}$ under the function
$\Phi$,
\[
L_{\pm}\equiv
\Phi^{-1}\left(\left\{\pm1/\sqrt{\kappa\,\zeta}\right\}\right),
\]
are nonempty, they are composed of a countable union of many
1--dimensional, 2--dimensional and 3--dimensional analytical
submanifolds of $\langle \!\langle \hbox{${\cal J}$\kern -.645em
{\raise.57ex\hbox{$\scriptscriptstyle(\ $}}}\rangle\!\rangle$. At
first sight, 0--dimensional (point--like) submanifolds are also
admissible, but in our case they are excluded by the stationarity.
For a proof of the quoted results in ${\rm{I\!R}}^3$ see
e.g.~\cite{MullerHRS73}, the extension to ${\rm{I\!R}}^4$ does not
present any problem.

A direct implication of these results is that in principle the
equality (\ref{eq:closed}) is valid just in $\langle \!\langle
\hbox{${\cal J}$\kern -.645em
{\raise.57ex\hbox{$\scriptscriptstyle(\ $}}}\rangle\!\rangle
\!\setminus\!\left(L_{+}\cup L_{-}\right)$, but by the continuity
of the left hand side of (\ref{eq:closed}) in the whole of
$\langle \!\langle \hbox{${\cal J}$\kern -.645em
{\raise.57ex\hbox{$\scriptscriptstyle(\ $}}}\rangle\!\rangle$, and
in particular through the lower dimensional surfaces that
constitute $L_{\pm}$, the left hand side of (\ref{eq:closed})
vanishes also in $L_{\pm}$.

Provided that expression (\ref{eq:closed}) is valid in all the
domain of outer communications $\langle \!\langle \hbox{${\cal
J}$\kern -.645em {\raise.57ex\hbox{$\scriptscriptstyle(\
$}}}\rangle \!\rangle $, it follows from the simple connectedness
of this region~\cite{ChrWald95}, and the well--known Poincar\'e
lemma, the existence of a global potential $U$ in the whole of
$\langle \!\langle \hbox{${\cal J}$\kern -.645em
{\raise.57ex\hbox{$\scriptscriptstyle(\ $}}}\rangle \!\rangle $
such that
\begin{equation}
\left(1-\kappa\,\zeta\,\Phi^2\right)^{2/3}
\mbox{\boldmath$\omega$}=\mbox{\boldmath$dU$}.
\label{eq:explicit}
\end{equation}

The previous potential $U$ is constant in each connected component
of the event horizon ${\cal{H}}^{+}$. This follows from the fact
that on the one hand $\mbox{\boldmath$\omega$}=0$ in
${\cal{H}}^{+}$, since this region is a Killing horizon whose
normal vector coincides with {\boldmath$k$}, and on the other
hand, as it was previously mentioned (see~\cite{Hawking72a,H-E}),
in order to that ${\cal{H}}^{+}$ be a Killing horizon the scalar
field must be analytical, especially at the horizon, hence
Eq.~(\ref{eq:explicit}) implies that $U$ is constant in each
connected component of the horizon.

The same result is achieved as well for the asymptotic regions,
because any stationary black hole with a bifurcate Killing horizon
admits a maximal hypersurface asymptotically orthogonal to the
stationary Killing field {\boldmath$k$}~\cite{ChrWald94}.
Furthermore, it has been shown~\cite{RaczWald96} that a stationary
black hole can be globally extended to other enlarged one
possessing a bifurcate Killing horizon.

In what follows we will show that besides the fact that the
potential $U$ is constant at the horizon and the asymptotic
regions, it will be also constant in the whole of
$\langle\!\langle \hbox{${\cal J}$\kern -.645em
{\raise.57ex\hbox{$\scriptscriptstyle(\ $}}}\rangle \!\rangle $.
For the case of minimal coupling ($\zeta=0$) this results implies
directly the staticity (\ref{eq:explicit}) as it has been
previously proved by Heusler~\cite{Heusler93}. We will extend his
proof to the case of non--minimally coupled to gravity scalar
fields.

For every function $f$ and 1--form {\boldmath$\Omega$}, the
following identity is satisfied (see the appendix
in~\cite{HeuStrau} for the details of the remaining calculations)
\[
\mbox{\boldmath$d^{\,\dagger}$}
\left(f\mbox{\boldmath$\Omega$}\right)=
f\mbox{\boldmath$d^{\,\dagger}\Omega$}
-\left(\mbox{\boldmath$df$}|\mbox{\boldmath$\Omega$}\right),
\]
here $\mbox{\boldmath$d^{\,\dagger}$}=\mbox{\boldmath$*\,{d}\,*$}$
stands for the co--differential operator. Applying this expression
to the function $U$ and the 1--form
$\mbox{\boldmath$\Omega$}/V^2$, together with (\ref{eq:explicit}),
the following can be obtained
\begin{equation}
\mbox{\boldmath$d^{\,\dagger}$}
\left( U\frac{\mbox{\boldmath$\omega$}}{V^2} \right)=
-\frac{\left(1-\kappa\,\zeta\,\Phi^2\right)^{2/3}
\left(\mbox{\boldmath$\omega$}
|\mbox{\boldmath$\omega$}\right)}{V^2},
\label{eq:dU2}
\end{equation}
where the forthcoming identity has been used (the proof of it can
be seen in Ref.~\cite{HeuStrau})
\[
\mbox{\boldmath$d^{\,\dagger}$}
\left(\frac{\mbox{\boldmath$\omega$}}{V^2}\right)=0.
\]

Following the same procedure used by Heusler
in~\cite{Heusler93,HeuStrau} we will integrate (\ref{eq:dU2}) over
a spacelike hypersurface $\Sigma$, with volume form
{\boldmath$i_{\mbox{\boldmath$k$}}\eta$}. Taking in consideration
that for any stationary 1--form {\boldmath$\Omega$}
($\mbox{\boldmath${\pounds}_k\Omega$}=0$)
\[
\mbox{\boldmath$d^{\,\dagger}\Omega\,
i_{\mbox{\boldmath$k$}}\eta$}=
-\mbox{\boldmath${d}\,*$}\left(\mbox{\boldmath$k$}\wedge
\mbox{\boldmath$\Omega$}\right),
\]
holds~\cite{HeuStrau}, the following relation is obtained applying
the Stokes theorem to (\ref{eq:dU2}),
\begin{equation}
\int_{\partial\Sigma}U{\mbox{\boldmath$*$}
\left(\mbox{\boldmath$k$}\wedge
\frac{\mbox{\boldmath$\omega$}}{V^2}\right)}=
\int_\Sigma\frac{\left(1-\kappa\,\zeta\,\Phi^2\right)^{2/3}
\left(\mbox{\boldmath$\omega$}
|\mbox{\boldmath$\omega$}\right)}{V^2}\,
\mbox{\boldmath$i_{\mbox{\boldmath$k$}}\eta$}.
\label{eq:int}
\end{equation}

Using now the identity~\cite{HeuStrau}
\[
2{\mbox{\boldmath$*$}\left(\mbox{\boldmath$k$}\wedge
\frac{\mbox{\boldmath$\omega$}}{V^2}\right)}=
\mbox{\boldmath$d$}\left(\frac{\mbox{\boldmath$k$}}{V}\right),
\]
the expression (\ref{eq:int}) can be brought to the form
\begin{equation}
\frac12\int_{\partial\Sigma}U
\mbox{\boldmath$d$}\left(\frac{\mbox{\boldmath$k$}}V\right)=
\int_\Sigma\frac{\left(1-\kappa\,\zeta\,\Phi^2\right)^{2/3}
\left(\mbox{\boldmath$\omega$}|\mbox{\boldmath$\omega$}\right)}
{V^2}\,
\mbox{\boldmath$i_{\mbox{\boldmath$k$}}\eta$}.
\label{eq:zero}
\end{equation}

The boundary $\partial\Sigma$ is constituted at its interior by
the event horizon ${\cal{H}}^{+}\cap\Sigma$, and at the infinity
by the asymptotic regions. Since the potential $U$ is constant
over each one of the connected components of these boundaries, it
can be pulled out from each one of the corresponding boundary
integrals in the left hand side of (\ref{eq:zero}).

The asymptotic regions and the connected components of the horizon
are all topological 2--spheres~\cite{H-E,ChrWald95}, by this
reason the left hand side of (\ref{eq:zero}) vanishes; from Stokes
theorem, the integral of an exact form over a manifold without
boundary is zero. Thereby, it is satisfied that
\begin{equation}
\int_\Sigma\frac{\left(1-\kappa\,\zeta\,\Phi^2\right)^{2/3}
\left(\mbox{\boldmath$\omega$}
|\mbox{\boldmath$\omega$}\right)}{V^2}\,
\mbox{\boldmath$i_{\mbox{\boldmath$k$}}\eta$}=0.
\label{eq:final}
\end{equation}

The integrand in (\ref{eq:final}) is non--negative, due to the
fact that {\boldmath$\omega$} is a spacelike 1--form since it is
orthogonal by definition to the timelike field {\boldmath$k$}
(\ref{eq:tor}). Hence, it follows that (\ref{eq:final}) is
satisfied if and only if the integrand vanishes in $\Sigma$, and
by stationarity it also vanishes in all the domain of outer
communications $\langle \!\langle \hbox{${\cal J}$\kern -.645em
{\raise.57ex\hbox{$\scriptscriptstyle(\ $}}}\rangle \!\rangle $,
consequently
\begin{equation}
\left(1-\kappa\,\zeta\,\Phi^2\right)^{2/3}
\left(\mbox{\boldmath$\omega$}
|\mbox{\boldmath$\omega$}\right)=0.
\label{eq:alter}
\end{equation}

From the previous conclusion (\ref{eq:alter}), it follows that
$\mbox{\boldmath$\omega$}=0$ in $\langle \!\langle \hbox{${\cal
J}$\kern -.645em {\raise.57ex\hbox{$\scriptscriptstyle(\
$}}}\rangle\!\rangle \!\setminus\!\left(L_{+}\cup L_{-}\right)$,
but by the continuity of {\boldmath$\omega$} in all of $\langle
\!\langle \hbox{${\cal J}$\kern -.645em
{\raise.57ex\hbox{$\scriptscriptstyle(\ $}}}\rangle\!\rangle$, and
in particular through the lower dimensional surfaces that
constitute the regions $L_{\pm}$, {\boldmath$\omega$} vanishes
also in $L_{\pm}$ and accordingly in all the domain of outer
communications $\langle \!\langle \hbox{${\cal J}$\kern -.645em
{\raise.57ex\hbox{$\scriptscriptstyle(\ $}}}\rangle \!\rangle $.
Hence, the staticity theorem is proved.

\section{Conclusions}

Finally, it is concluded that for a non--rotating strictly
stationary black hole with a self--interacting scalar field
non--minimally coupled to gravity, the corresponding domain of
outer communications is static if analytic field configurations
are considered. As in the minimal case~\cite{Heusler93} this
result remains valid when no horizon is present.

\section{Acknowledgments}

The author thanks Alberto Garc\'{\i}a by its incentive in the
study of these topics and by its support in the course of the
investigation. This research was partially supported by the
CONACyT Grant 32138E. The author also thanks all the encouragement
and guide provided by his recently late father: Erasmo Ay\'on
Alayo, {\em Ibae Ibae Ibayen Torun}.

\end{document}